\documentstyle[12pt]{article}

\author{Reza Tavakol$^\flat$\thanks{E-mail address: reza@maths.qmw.ac.uk}
\ and Roustam Zalaletdinov$^{\flat\sharp}$\thanks
{E-mail address: rmz@maths.qmw.ac.uk; 
address after 10 May 1996: N.Copernicus Astronomical Center,
Polish Academy of Sciences, ul. Bartycka 18, 
00-716 Warsaw, Poland; 
e-mail address: zala@camk.edu.pl}\\[5mm]
{\em $^\flat$School of Mathematical Sciences,
Queen Mary and Westfield College}\\
{\em University of London, Mile End Road, London E1 4NS, England, U.K.}\\[2mm]
{\em $^\sharp$Department of Theoretical Physics,
Institute of Nuclear Physics}\\
{\em Uzbek Academy of Sciences, Ulugbek, Tashkent 702132, Uzbekistan, C.I.S.}}
\title{\LARGE \bf On the Domain of Applicability of \\
General Relativity
}
\date{}
\def\be{\begin{equation}}
\def\ee{\end{equation}}
\def\bea{\begin{eqnarray}}
\def\eea{\end{eqnarray}}

\vspace{0.6cm}
\normalsize
\vspace{2cm}
\vspace{2cm}

\begin{document}

\maketitle
\begin{abstract}
We consider the domain of applicability of general relativity (GR), as a
classical theory of gravity, by considering its applications to a variety of
settings of physical interest as well as its relationship with real
observations. We argue that, as it stands, GR is deficient whether it is
treated as a microscopic or a macroscopic theory of gravity. We briefly
discuss some recent attempts at removing this shortcoming through the
construction of a macroscopic theory of gravity. We point out that such
macroscopic extensions of GR are likely to be non-unique and involve
non-Riemannian geometrical frameworks.
\end{abstract}

\sloppy

\sloppy

\sloppy

\sloppy

\thispagestyle{empty}

\newpage

\section{Introduction}

Ever since the inception of general relativity (GR), many attempts have been
made to extend or generalise it. The motivation for these attempts has
mainly come from outside, relating essentially to the questions of its
quantisation and its unification with other forces of nature. They therefore
amount to an {\it external} critique of GR. In its applications to classical
gravitational phenomena, however, GR has been generally assumed to be
satisfactory. \\

Our aim here is to take GR as a classical theory of gravity and ask: \\

(i) what is its precise domain of applicability? \\

(ii) can it unambiguously deal with real settings and observations? \\

In this sense we are concerned with an {\it internal} critique of the
theory. \\

We start by recalling that both in testing and applying GR as a theory of
gravity, it is usually assumed\footnote{There are, however, exceptions. See,
for example, Ellis
\cite{Elli:1984}, Sciama \cite{Scia:1971}, Ehlers \cite{Ehle:1973}.} that
both the gravitational phenomena under study and the observations involved
in deducing information about them are {\it ideal}, in the sense that
particles are taken to be {\it test}\footnote{By a test (or ideal)
particle we mean a non-rotating freely falling
point particle \cite{EPS:1972}.} particles 
and the observations are assumed to have {\it 
infinite} resolution. Clearly, in {\it real} applications, neither of these
assumptions could be justified. The question then arises as to the precise
status of GR when applied to real phenomena and put into correspondence with
real observations. \\

To answer this, we discuss some of the major difficulties that arise when GR
is applied in real settings. It turns out that the central problem is
naturally tied up with the fact that both real observations and real
phenomena are {\it extended} in nature; the former due to the fact that all
real observations unavoidably involve finite resolutions and the latter
because, strictly speaking, there are no point particles in reality. To
demonstrate the difficulty, we consider two problems: the motion of
particles and the question of cosmology. These considerations naturally lead
us to the question of whether GR can be consistently treated either as a 
{\it microscopic} or {\it macroscopic} theory of gravity. We shall discuss
each of these scenarios in turn and argue that, as it stands, GR is
deficient in both cases. We shall then briefly consider the question of
whether the theory can be successfully made compatible with real
observations and phenomena {\it internally} (i.e by only employing concepts
internal to GR) and {\it uniquely}. This leads us to the question of
averaging and macroscopic considerations of GR which we shall briefly
discuss and conclude that the answer is again likely to be negative on both
accounts. \\

We should also add here that despite our employment of GR as {\it the}
classical theory of gravity, the main points raised here are also of
relevance for other (alternative) theories of gravity, and in their
comparison with observations. As a result what we have to say is of
potential importance in making a comparative study of such theories and
therefore in determining the "correct" classical theory of gravity. \\

The organisation of the paper is as follows. In Section 2 we consider the
application of GR to cosmology and the problem of motion of bodies. In
Section 3 we discuss the correspondence of GR with observations and discuss
some of the problems that arise when comparison is made between predictions
of GR and observations. A statement of the problem of the domain of
applicability of GR is discussed in Section 4. Sections 5 and 6 contain the
treatment of GR respectively as microscopic and macroscopic theories of
gravity, together with some of the features and shortcomings of each
scenario. In Section 7 we briefly discuss some of the attempts that have
recently been made to develop a macroscopic theory of gravity, together with
their corresponding problems and finally Section 8 contains our conclusions. 

\section{Applications of GR in some physical settings}

The usual starting point in the classical studies of gravitational
phenomena, including the universe itself, are Einstein's field equations
(EFE) 
\begin{equation}
\label{efe}r_{ab}-\frac 12g_{ab}g^{cd}r_{cd}=-\kappa t_{ab}, 
\end{equation}
together with the equations of motion 
\begin{equation}
\label{em}t_{a;b}^b=0, 
\end{equation}
which are known to follow from (\ref{efe}). It is, therefore, these
equations which are employed in order to interpret local (e.g. solar system)
and large scale (e.g. cosmological) observations, on the one hand, and to
construct mathematical models in order to predict the evolution of
gravitational phenomena, on the other. \\

From a theoretical point of view, the main idea is to treat these equations
as a correspondence rule\footnote{We 
shall not dwell on the exact nature
of this correspondence, specially its (i) {\it uniqueness}, i.e
whether given $t_{ab}$, the metric $g_{ab}$ is specified uniquely
and (ii) 
{\it stability},
i.e. whether small errors in the specification of $t_{ab}$
or the simplifying assumptions usually employed in the construction 
of models can give rise to qualitative changes in the 
corresponding geometry \cite{rtge88,acrt92}. These questions are, however, of
potential importance in determining the overall status of the 
theory in {\it practice}.}, whereby given the form of the stress-energy
tensor $t_{ab}$, the geometry can be specified by solving 
equations (1)\footnote{In reality one assumes, in addition
to $t_{ab}$, a number of simplifying assumptions, such as
symmetry, which partially nail down the geometry as well, as is the 
case, for example, in cosmology,
with the assumptions of isotropy and homogeneity
which lead to 
Fried\-mann--Le\-ma\^{\i}tre--Ro\-bert\-son--Wal\-ker (FLRW)
geometry. Einstein's field equations will then specify the 
unknown function(s) in the metric and
hence the dynamical evolution of the geometry. 
This is what Ellis \cite{Elli:1995} refers
to as the {\it direct} method in cosmology.}. The resulting predictions of
the theory are then to be compared with real observations, in order to
ascertain its viability as a theory of gravity on all relevant scales,
including the cosmological ones. \\

Now it is well known that general relativity has been extremely successful
in accounting for local observations, including the usual classical tests in
the solar system and the observations of binaries \cite{Will:1993}. We shall
come back to the question of motion of bodies in GR in the following
sections. Here, however, it is instructive to contrast this with the status
of GR as a theory of gravity on cosmological scales, for which there is
little detailed direct evidence. The main reason for this is not just the
usual ``uncertainty'' brought about by the error bars which are bound to be
present in all cosmological observations. It relates also to the very nature
of real cosmological observations and the difficulties that arise when
attempts are made to construct an appropriate theoretical framework for
their interpretations within GR.

To discuss the domain of applicability of GR as a classical theory of
gravity we shall focus on a number of settings of physical importance -
namely the cases of cosmology and the motion of particles - where
applications of GR involve fundamental difficulties. 

\subsection{GR and cosmology}

A fundamental problem arising in the applications of GR to 
cosmology\footnote{As well as all other alternative theories of
gravity proposed so far.} concerns the question of scales over which the
theory is supposed to hold. This is due to the presence of a hierarchy of
the cosmological scales of observational interest in the 
universe\footnote{For example, Ellis \cite{Elli:1984}
singles out five such scales, down to the scale 
of relevance for stars.}, on the one hand, and the absence of an intrinsic
scale in the theory, on the other.

In the usual practice of standard cosmology, this problem is circumvented by
making the following assumptions, which are usually made 
implicitly\footnote{There
are authors, however, who mention this
assumption explicitly. See, 
for example, \cite{Elli:1984,Shir-Fish:1962,Hemm:1987,Zala:1992}.} : \\

\noindent (1) in the real complicated ''lumpy'' universe with a discrete
matter distribution (of stars, galaxies, clusters of galaxies, etc.), the
stress-energy tensor $t_{ab}=t_{ab}^{{\rm (discrete)}}$ can be adequately
approximated by a ''smoothed'', or hydrodynamic, stress-energy tensor $
T_{ab}^{{\rm (hydro)}}$, usually taken to be representable by a simple
perfect fluid. \\

\noindent (2) as $t_{ab}^{{\rm (discrete)}}\rightarrow T_{ab}^{{\rm (hydro)}}
$ on the right hand side of the EFE, the left hand side remains unchanged
under such a change and therefore the appropriate field equations for
describing the matter distribution $T_{ab}^{{\rm (hydro)}}$ still take the
form 
\begin{equation}
\label{efe:ss}R_{ab}-\frac 12G_{ab}G^{cd}R_{cd}=-\kappa T_{ab}^{{\rm (hydro)}
}, 
\end{equation}
where capital letters denote quantities which correspond to the smoothed
matter distribution $T_{ab}^{{\rm (hydro)}}$, and $G_{ab}$ and $R_{ab}$ are
the metric and the Ricci tensors describing the corresponding geometry which
is taken to be the same pseudo-Riemannian spacetime geometry as for (1). \\

\noindent (3) the corresponding equations of motion follow from (3) and are
given in the form 
\begin{equation}
\label{em:ss}T_{a;b}^{b{\rm (hydro)}{}}={}0, 
\end{equation}
with $T_a^{b{\rm (hydro)}{}}=T_{ac}^{{\rm (hydro)}}G^{cb}$. \\

Given the intricacy and the detail involved in the real stress-energy tensor 
$t_{ab}=t_{ab}^{{\rm (discrete)}}$, these assumptions allow a large number
of astrophysical and cosmological problems to be treated, which would
otherwise have remained impossibly difficult to tackle. Implicit in these
assumptions is that the solutions of (\ref{efe:ss}) approximate well the
solutions of (\ref{efe}) with the corresponding $t_{ab}^{{\rm (discrete)}}$,
at least in the regions between the discrete matter constituents. This is a
very strong theoretical assumption which is usually made without
justification, in applications of GR to cosmological 
problems\footnote{We shall
see what this assumption 
implies in Section 7.}. The main point here is that it is not a priori clear
how good these assumptions are. What is therefore required, is a consistent
application of GR to cosmology, where both sides of the EFE (\ref{efe}) are
averaged out simultaneously, in order to find out the extent to which the
above assumptions are in fact justified (see \cite
{Elli:1984,Shir-Fish:1962,Zala:1992,Kras:1996} and references therein). 

\subsection{Motion of bodies in GR}

Another important problem of real significance, from a physical point of
view, is the description of motion of particles. There are two ways in which
particles are treated within the context of GR: either as ideal particles or
as what we shall refer to as {\it real}\footnote{By real or
ordinary particles, in the language of
Ehlers \cite{Ehle:1980}, we mean extended, deformable
(usually spinning) structures
representable as connected spatially compact sub-manifolds with
timelike boundary hypersurfaces. We note in passing
that since there are no
rigid bodies in GR, any extended body would have
an infinite number of degrees of freedom.} particles. We shall briefly
discuss each of these in turn. \\


\subsubsection{Ideal particles}


The usual notion of particle motion in GR concerns the motion of ideal
(test) particles, which are postulated to move along timelike geodesics of
the (pseudo)-Riemannian spacetime manifold. The problem, however, is that so
far no unambiguous and problem-free model has been put forward in the
context of GR, for the description of the motion of such particles. The main
difficulty stems from the problematic relation between the field equations
and the equations of motion. It is in fact important to distinguish between
the ``laws of motion'' and the ``equations of motion'' (see \cite
{Ehle:1980,Hava:1979,Hava-Gold:1962}). The main issue is that even though
equations (2) follow directly from (1), only in certain special
circumstances, such as the case of monopole singularities, where the
influence of a moving monopole particle on the background metric is
negligible, does the geodesic law of motion follow 
from (2)\footnote{For a more precise 
formulation of the relationship between the EFE and the equations
of motion see Ehlers \cite{Ehle:1987}.}. The difficulty arises from the fact
that even though one can construct model candidates for point-like particles
which satisfy equations (2), no such examples exist as yet which also
satisfy the field equations. For example, it is known \cite
{Ehle:1987,Scha:1984} that given a smooth spacetime, a
distributional model of point-particle energy-momentum tensor $t^{ab{\rm 
(particle)}}$ supported by a timelike worldline $L=\zeta (s)$, which obeys
the dominant energy condition and satisfies (2), is necessarily monopolar 
\begin{equation}
t^{ab{\rm (particle)}}(x)={\rm const}\int \dot \zeta ^a\dot \zeta ^b\delta
(x-\zeta (s))ds,
\end{equation}
where ${\rm const\ }>0$ and $L$ is a geodesic. The problem is that this is
not compatible with the EFE and worse still even the field equations
themselves lose their mathematical validity, since the metric diverges at
the particle location. \\

On the other hand, modelling of ideal particles moving along geodesics as an
exact solution to EFE representing a test particle moving along a geodesic
has not been successful as yet (see \cite{Infe-Schi:1949,Nevi:1995} and
references therein).\\ 

An alternative approach to the problem of ideal particles is to treat them
as extended bodies\footnote{Represented as timelike world-tubes,
with finite spacelike 3-volume 
sections, which are the supports of the corresponding interior 
stress-energy tensors $
T_{ab}^{{\rm (interior)}}$, describing the interior of the bodies 
and vanishing
outside \cite{Papa:1974,Dixo:1979}.}, in the limit of all moments
characterising them vanishing. Such point-like, {\em single-pole}, extended
bodies appear to avoid the difficulties with the infinite matter density and
metric divergences. The problem, however, is that, similar to the previous
case, there is no way of combining the equations of motion (2) with the
field equations (\ref{efe}) 
(with $t_{ab}=T_{ab}^{{\rm (interior)}}$)\footnote{We should also add here 
that the derivation of
the geodesic equations for a point particle, based on the variational
principle given by Fock \cite{Fock:1959}, uses essentially a similar
representation of a point particle as a stress-energy tensor over a timelike world
tube with the subsequent limit of a concentrated mass.}. \\

\subsubsection{Real particles}

Turning now to the motion of extended particles, we recall that despite a
great deal of efforts, as yet no non-approximate full treatment of motion of
real particles exists \cite{Papa:1974,Dixo:1979,Damo:1983}. The difficulties
are similar to those encountered in a realistic treatment of cosmology,
including the requirement for coordinate free definitions of integrals of
tensor fields over the spacelike surfaces of the particles' world-tubes, in
order to determine moments in terms of which such particles could be
characterised. A great deal of effort has gone into deriving approximate
equations of motion of such particles, up to orders where gravitational
backreaction also appears \cite{Damo:1983}, but it still remains unclear
whether such formal approximations converge or provide metrics that
approximate actual solutions \cite{Ehle:1987}. \\

Another important feature of the motion of real particles is their
non(Riemannian)-geodesic nature. To get a partial feel for this, let us
consider the motion of a small spinning sphere of radius $R$, which is
assumed to have zero quadruple and higher multipoles \cite
{Papa:1974,Dixo:1979}. The motion of the particle can then be represented by
a line $L$ inside its world tube, the points of which are denoted by $X^i$.
Assuming the particle to be small, letting $R\rightarrow 0$ and $\delta
x^i=x^i-X^i$, the equation of motion of such a particle may be written as 
\begin{equation}
\label{motion}\frac D{Ds}\left( mu^i\right) +\frac D{Ds}\left( u_j\frac{
DS^{ij}}{Ds}\right) +\frac 12S^{kl}u^m{R^i}_{mkl}=0,
\end{equation}
where the $S^{kl}$ take account of the particle's spin and are defined as 
\begin{equation}
S^{ij}=\int \delta x^iT^{j4{\rm (interior)}}dv-\int \delta x^jT^{i4{\rm 
(interior)}}dv.
\end{equation}

This shows clearly that the first term in equation (7), $D\left( mu^i\right)
/Ds=0$, corresponds to the usual Riemannian geodesic equation in the $g_{ij}$
space, with the other terms appearing as soon as the $S^{kl}$'s are
non-zero, i.e. the internal degrees of freedom of particles - here the spin
- are taken into account. These extra spin-dependent terms clearly
demonstrate that the motion of such particles cannot be geometrised within
the Riemannian geometry and therefore a more general geometrical framework
is required for their geometrisation. \\

Another important difficulty arises when one considers real observations,
which we shall turn to now. 

\section{The correspondence between theory and observations in GR}

In order to test the viability of physical theories (including GR),
especially in their role as theoretical frameworks within which observations
are analysed, it is necessary to devise ways whereby they can be put into
one-to-one correspondence with observations and measurements. The setting up
of such a correspondence would involve two steps: (i) to locate quantities
within each theory which have observational counterparts or can be expressed
in terms of such quantities and (ii) to ensure that such quantities possess
identical spacetime domains of definition as those employed in observations.
We shall refer to theories characterisable in terms of such {\em 
operationally defined} quantities as {\em complete}. An important feature of
completeness is that it makes the question of viability of the theories
decidable\footnote{We should emphasise the
distinction between the viability of a theory 
and its completeness. Completeness is a necessary but
not a sufficient condition for the viability of 
a theory.}. \\

Now the first of these steps is not in principle difficult, even though
there are quantities in physical theories which are not operationally
defined, or inversely, there are physically motivated concepts which cannot
be unambiguously defined in certain theories, such as, for example, the
notion of energy and mass in GR. Regarding the second step, the main
difficulty in most (dynamical) physical theories arises from the fact that
they are formulated as differential equations, whereas all observational
devices have a finite resolution. This is of vital importance, especially in
view of the fact that, as was emphasized long ago by Bohr and Rosenfeld \cite
{Bohr-Rose:1933}, only spacetime averages of field quantities have physical
meaning. Consequently, point-like quantities are not operationally
definable. This dichotomy is crucial in the case of GR, where the
mathematical quantities of the theory are essentially point-like whereas the
observations (especially in the case of cosmology) are invariably extended,
in the sense of covering a (large) neighbourhood in the spacetime. For
example, the theoretical procedures \cite{Ehle:1973} of measuring the
curvature tensor by means of the equations of geodesic deviation or through
the employment of geodesic triangles are infinitesimal in nature, as they
are based on standard calculus and therefore employ infinitesimal distances
and times. On the other hand, real measurements (and observations) are {\em 
extended} by nature, as they involve averages over spacetime regions of
finite characteristic length (see \cite{Bohr-Rose:1933,DeWi:1979}). \\

As an example of the fundamental difficulties of this type that one
encounters in GR, let us recall the constructive-axiomatic approach to GR
developed by Ehlers, Pirani and Schild (EPS) \cite{Ehle:1973,EPS:1972}. The
main aim of this work is to make transparent, in an axiomatic way, the
relationship between the geometrical structures, including the Riemannian
nature of spacetime, on the one hand, and the observable phenomena, on the
other. More precisely, starting from the paths of light rays and the
trajectories of idealised particles and relating them to the conformal and
projective structures of spacetime respectively, they claim that these
substructures, together with an additional assumption regarding the
constancy of norm of vectors under parallel transport, would uniquely fix
the underlying geometry of spacetime to be Riemannian. \\

There are two important points to note about this scheme: one theoretical
and one observational. Firstly, the EPS assumptions do not necessarily
reduce the geometry of spacetime to be Riemannian; this is only true if the
starting geometrical framework chosen is Weylian. Therefore, starting from
more general geometrical frameworks, such as for example the Finslerian one,
the imposition of EPS conditions does not necessarily reduce the underlying
geometry of the spacetime to be Riemannian \cite{Tava-VdBe:1986,Roxb:1991}.
In this sense then the whole scheme is dependent upon the theoretical
framework chosen, i.e. the starting Weylian geometrical framework. Secondly,
the type of observations implied by the EPS scheme are {\it ideal}, since in
this scheme it is the ideal particles which are used in order to determine
the projective structure of the spacetime. Now given that all real particles
are extended (and usually have spin), an important question, from the point
of view of our discussion here, is what happens if the idealised test
particles in the EPS scheme are replaced by real physical extended
(spinning) particles? In other words what is the resulting geometry for
which the trajectories of such particles are geodesics? This is a very
difficult problem to treat in its fullness, as even a satisfactory
description of motion of extended bodies does not as yet exit (see 
Dixon \cite{Dixo:1979} and Ehlers \cite{Ehle:1980,Ehle:1979} for a detailed
discussion). Nevertheless, the equation of motion of the spinning particle
discussed in the previous section indicates that GR (or at least its
Riemannian geometrical component) can not be made constructively
(-axiomatically) compatible with the motion of real (spinning) particles,
treated as primary objects of the theory. Therefore, the consideration of
real particles is likely to lead to geometrical settings that are more
general than Riemannian. We shall see this thread appearing again when we
consider the possibility of a macroscopic theory of gravity in Section 7. 

\section{Domain of applicability of GR}

Our considerations so far seem to indicate that GR has fundamental
difficulties in successfully treating both ideal and real particles, as well
as dealing with the cosmological phenomena. The main sources of these
difficulties (apart from the fact that no successful treatment of point
particles exists) are related to the facts that (i) real phenomena are
extended and (ii) real observations have finite resolutions and therefore
always involve some form of spacetime ``smoothing'' or ``averaging'' in
practice. \\

To see whether GR is complete, we start by asking whether there exist scales
over which Einstein's equations hold 
exactly\footnote{It is worthwhile to recall that this
question has an old history going back to
Einstein himself who realised the potential mathematical and physical
problems that arise when attempts are made to construct point-like matter
models within GR. This led Einstein to conclude that GR is adequate to
describe macroscopic processes with continuously distributed matter model
(see \cite{Ehle:1961} for a discussion).}. There are in principle two
different possible answers to this question: either GR, as it stands, is a 
{\em microscopic} theory or it is {\em macroscopic}, in the sense of being
adequate to describe gravitation on a specified range of scales, with given
matter models specifying these scales. We should contrast this with the
usual practice which employs GR in order to describe classical gravitational
phenomena on all scales, encountered both in theory and practice. \\

In the following Sections we consider each of these scenarios in turn and
ask whether GR can be successfully treated in either way. 

\section{GR as a microscopic theory}

In this section we start by assuming that GR, as it stands, is a classical
microscopic theory of gravity and ask whether we could do so consistently.
The usual motivation for treating GR as a microscopic theory of gravity
seems to be based on its success in accounting for local phenomena,
including the solar system tests and the observations of binaries. From a
more formal point of view, the approach of EPS, which sets up a
correspondence between the motion of ideal particles and the projective
geometry of spacetime, could also be counted as a plus for this
interpretation. There are, however, many problems with a microscopic
interpretation of GR. From both a physical and mathematical point of view,
the problems which arise when GR is taken as a classical microscopic theory
of gravity have similarities with those which arise when electrodynamics and
Newtonian gravity are considered as microscopic theories. There are
important differences, however, due to the specific nature of general
relativistic gravitation. The following are some of the defining features
and difficulties of such a microscopic interpretation. \\

{\bf (I) The energy-momentum tensor}: It assumed to be expressible as a
discrete distribution of point-like matter constituents (''particles'')
localized at points $x_A$, with the energy-momentum tensor supported by
timelike world lines and given by 
\begin{equation}
\label{em:micro}t_{ab}^{{\rm (micro)}}(x_1,x_2,...,x_N)=\sum_{A=1}^Nt_{ab}^{
{\rm (particle)}}(x_A),
\end{equation}
where $N$ is the number of the constituents\footnote{It is worth 
pointing out that $N$ is not 
sufficiently large here to make it physically adequate 
to apply statistical or kinetic methods for the description 
of the matter distribution (see \cite{Isra:1972} and references therein).}.
The corresponding field dynamics and the laws of motion are then given by (
\ref{efe}) and (\ref{em}) respectively, where in both cases the
stress-energy tensor is replaced by $t_{ab}^{{\rm (micro)}}$. \\

{\bf (II) Vacuum nature}: outside point sources, the general relativistic
microscopic gravitation is an inherently vacuum phenomenon\footnote{This is
physically similar to the case of
classical microscopic electrodynamics, as developed in the
Lorentz theory of electrons \cite{Lore:1916}, 
which describes the dynamics
of point-like charges in vacuum by means of 
microscopic field equations,
the Lorentz-Maxwell equations.}, satisfying the vacuum field equations 
\begin{equation}
\label{efe:micro} r_{ab}-\frac 12g_{ab}g^{cd}r_{cd}=0, 
\end{equation}
with the discrete sources (\ref{em:micro}) acting as some sort of boundary
conditions. The totality of such sources can be {\it defined} as the set of
all possible singularities of the solutions of the vacuum field equations (
\ref{efe:micro}), satisfying all possible symmetries, asymptotic 
conditions\footnote{The same
approach may be formulated aslo for classical microscopic
electrodynamics.}, etc. So, for example, an isolated point mass can be
defined as the singularity of the corresponding spherically symmetric static
solution of (\ref{efe:micro}) with the integration constant identified with
its mass. The task of compiling a complete list of all such microscopic
sources allowed by the equations (\ref{efe}) with (\ref{em:micro}), or (\ref
{efe:micro}), is extremely difficult and is equivalent to finding the set of
all vacuum solutions to (\ref{efe:micro}). As examples of candidates for
such sources we may consider, ideal particles, isolated point masses (i.e.
sources for Schwarzschild black holes), point masses with spin (i.e
intrinsic angular momentum), like the source for the Kerr 
solution\footnote{Though 
the structure of the Kerr solution 
singularity is known 
to have topologically
the structure of a singular ring, 
the Kerr solution has
essentially a vacuum nature,
and can therefore be considered as being of microscopic nature.} and
rotating (test or interacting) particles moving freely or in the field of
the other masses. There are also other examples of solutions to the vacuum
EFE \cite{KSMH:1980}, which are regular everywhere except at a number of
singularities and which could be thought of as special mass distributions,
such as the sources of Weyl's and Curzon's solutions, and solutions
involving more than one Schwarzschild or Kerr black holes. Such sources,
though ``extended'' in the sense of being non-local (line-like, etc.), can
still be considered as microscopic. \\

{\bf (III) Modelling point particles}: As was pointed out in Subsection
2.2.1, there are fundamental difficulties in successfully modelling point
particles within GR. One could, however, assume that such particles are {\it 
external} to the theory, as is the case with point-like charges in the
classical theory of electromagnetism \cite{Lore:1916}, and point masses in
Newtonian gravity. Regarding the former, we note that all attempts to
overcome the difficulties involved in modelling electrons, including the
infinities that arise due to the self-energy, failed until it was realised
that point charges must be treated as the singularities of the
electromagnetic field and they are therefore incompatible with the field
equations. In this case there are essentially two ways out, both external to
the original theory. The first involves quantum electrodynamics, where the
interaction of electrons are viewed quantum mechanically and where the
problem of self-energy of electron is dealt with by employing
renormalisations, and the second involves the employment of a continuous
model of charged matter, resulting in Maxwell's macroscopic electrodynamics,
which has been shown \cite{Lore:1916} to result from a spacetime smoothing
(or averaging) of the equations and relations of the microscopic theory. A
similar situation also arises if Newtonian gravity \cite{MTW:1973} is
considered as a microscopic theory. The field theoretic structures of both
microscopic (taken as standard Newtonian gravity) and macroscopic (upon a
space averaging of Newtonian gravity) theories can be shown to be the same,
apart from the discrete matter distribution being replaced by a continuous
one \cite{Zala:unpub}, which constitutes the problem of construction of
continuous matter models in Newtonian gravity.\\

{\bf (IV) Equations of motion}: As was discussed in Section 2, no
satisfactory treatment of ideal particles exists which is at the same time
compatible with the EFE. \\

{\bf (V) Newtonian limit}: One may expect that the correspondence principle
between GR, as a microscopic theory, and Newtonian gravity as a
non-relativistic microscopic theory of gravity, should in principle provide
a limiting procedure for the field equations (\ref{efe}) (together with (\ref
{em:micro})). For the left hand side of (\ref{efe}), there is a well-known
procedure which reduces it to the Poisson equation. For the right hand side,
the problem is how to define the limiting case for the sources. This is not
clear, but the Israel-Carter-Robinson theorems \cite{Cart:1979} may be
considered to have established, at least for the case of one isolated mass
(Schwarzschild solution), that the simplest general relativistic analogues
of point-like sources in Newtonian gravity are black holes. As far as the
geodesics of the test particles are concerned, the correspondence holds
trivially. What has not been shown, however, is that the solution of the
field equations for a test particle \cite{Infe-Schi:1949,Nevi:1995}  (see
Section 2) possesses a Newtonian limit, nor has it been proved that the
Newtonian limit for the case of $N$ point-like sources (black holes, test
particles, etc.) exists (see \cite{Hava:1979,Ehle:1987} for references). It
therefore follows that there is as yet no rigorous formulation of the
correspondence principle for GR, treated as a microscopic 
theory\footnote{For example not all vacuum 
solutions of Einstein's equations are known to
have a Newtonian limit \cite{KSMH:1980,Ehle:1989}.}. \\

{\bf (VI) Relation to observations}: Because of the finiteness of their
resolutions, all observations (and measurements) involve finite regions of
spacetime and are therefore extended in the sense of involving averages
(time, space, spacetime, statistical, etc.) of measured quantities. On the
other hand, GR as a microscopic theory cannot {\it internally} produce
quantities which are similarly extended and are therefore operationally
definable. This is of relevance both for the case of extended particles as
well as for cosmology. \\

These considerations indicate that microscopic GR cannot be complete. We
should mention here that there are real settings, however, where it is
possible (and physically adequate) to approximate real matter sources by
point-like models. In such cases, we may treat GR as an adequate microscopic
theory {\it in practice}. We should, however, bear in mind that this is an
approximation and the important question in this respect is the estimation
of the errors that this approximation involves in each particular setting. 

\section{GR as a macroscopic theory}

Next we consider whether GR, as it stands, can be considered as a consistent
macroscopic theory of gravity. If so, a great deal of objections raised
above against the microscopic theory, such as compatibility with extended
phenomena would be removed. Further, since macroscopic theory presupposes
its objects to be averages, it would therefore be compatible with non-local
measurements and observations.\\

The following are some of the defining features and difficulties that arise
when GR is treated as a macroscopic theory: \\

{\bf (I) The energy-momentum tensor:} The main assumptions of such a theory
are that (a) there exist macroscopic (continuous) matter 
distributions\footnote{The macroscopic 
model of matter in a classical macroscopic theory of gravity
(assumed here to be GR) is to be postulated analogously to that
of the macroscopic model of charge and current distributions in
Maxwell's classical theory of electromagnetism \cite{Pano-Phil:1962}, where 
charge $\rho ^{{\rm (macro)}}$ and current
distributions ${\bf j}^{{\rm (macro)}}$ are given as continuous
(hydrodynamic) distributions
$\rho ^{{\rm (macro)}}(x)=\rho ^{{\rm (hydro)}}(x)$ and
${\bf j}^{{\rm (macro)}}(x)={\bf j}^{{\rm (hydro)}}(x)$ together
with the defintion of a charge configuration, its boundary and its
exterior.} with hydrodynamic stress-energy tensors $T_{ab}^{{\rm (macro)}}$, 
\begin{equation}
\label{em:macro}T_{ab}^{{\rm (macro)}}(x)=T_{ab}^{{\rm (hydro)}}(x),
\end{equation}
which are supported by world-tubes $\Sigma $ (a region of spacetime filled
with matter), $x\in \Sigma $, and (b) macroscopic gravitational field
dynamics due to macroscopic matter distributions (\ref{em:macro}) are
governed by Einstein's field equations (\ref{efe:ss}), which in turn define
an averaged metric $G_{ab}$. The corresponding law of motion is then given
by (\ref{em:ss}) and, unlike microscopic theory which essentially has a
vacuum character, the macroscopic theory is supposed to describe the field
both inside and outside extended 
bodies\footnote{This is similar to the case of classical
macroscopic electrodynamics (see, for example, \cite{Pano-Phil:1962}), 
which describes 
the electromagnetic field and its dynamics for the extended 
charge and current 
distributions inside the 
sources and electrovacuum fields outside the sources. 
The sources for macroscopic gravitational fields 
are extended configurations with
an interior region (matter), a boundary (a 3-surface which is the section of
the world-tube), and
an exterior (vacuum, or more precisely, macroscopic vacuum) region, which is 
to be distinguished from the microvacuum.}. \\

Mathematically, the continuous matter model (\ref{em:macro}) may take the
form of any stress-energy tensor supported by a world-tube and satisfying
the appropriate differentiability and energy 
conditions \cite{Hawk-Elli:1972}. From a physical 
point of view, however, in addition to details of
hydrodynamics and thermodynamics, one also requires information regarding
the scales over which physical quantities are defined. \\

{\bf (II) The nature of macroscopic gravity:} The macroscopic gravitational
field due to a matter configuration satisfies the field equations (\ref
{efe:ss}) inside of that configuration and the macroscopic vacuum equations 
\begin{equation}
\label{efe:macro} R_{ab}-\frac 12G_{ab}G^{cd}R_{cd}=0 
\end{equation}
in the exterior region. Also due to the matter characteristics undergoing a
discontinuous jump at the boundary of the body, the field characteristics
must be matched on the boundary in an appropriate way (see for example, \cite
{KSMH:1980}, and the references therein)\\

{\bf (III) Motion of real particles}: As was mentioned in Section 2, no
consistent description exists of motions of extended sources within GR,
which is at the same time compatible with the field equations. Of potential
interest here is the result by Lichnerowicz \cite{Lich:1967} which shows
that EFE together with a suitable source model (e.g., perfect fluid)
determines the motion of the source as well as the evolution of the
gravitational field. Interestingly, however, there is no analogous result
for a single extended body in vacuum, or a system of $N$ such bodies \cite
{Ehle:1987}. \\

{\bf (IV) Correspondence with microscopic theory}: Starting from GR as a
macroscopic theory, a fundamental question is the nature of the
corresponding underlying microscopic theory from which the macroscopic
theory may be derived upon some assumptions. \\

{\bf (V) Newtonian limit}: It is not clear what the correspondence principle
for such a theory is and what should be taken as the macroscopic analogue of
the Newtonian theory\footnote{ 
For Newtonian gravity treated as a macroscopic theory, one considers the
Poisson equation with matter distribution $\mu $$^{{\rm (macro)}}$ in
the form $\mu ^{{\rm (macro)}}(x)=\mu ^{{\rm (hydro)}}(x)$, together
with Newton's equation of motion for an element of the medium
with a test mass $\mu _0$ moving in the field created 
by the gravitational potential due to
the distribution $\mu $$^{{\rm (hydro)}}$.} in this case. \\

{\bf (VI) Intrinsic scales}: An important distinguishing feature of GR
(relative to electrodynamics and Newtonian theory) is the nonlinearity of
its field equations. An immediate consequence of this is that by assuming
the left hand side of the EFE  to remain of the same form as (3), one is
ignoring any reference to intrinsic scales implied in the definition of $
T_{ab}^{{\rm (macro)}}$ and, therefore, the corresponding correlations. It
should be emphasised that in such a theory it is $T_{ab}^{{\rm (macro)}}$
which carries all the information about scales and since its definition
relies totally on the model of matter, the effect of changes in scales only
concerns the right hand side and not the fundamental structure of the field
equations themselves. A related problem is whether there exists a built-in
scale which may serve as a correlation lenght in GR. This is unlikely, in
view of the fact that GR as a classical theory of gravity has only two
constants, namely the universal constant of gravitation $G$ and the speed of
light $c$, neither of which (nor any combination of them) possess such a
scale\footnote{Here we are confining
ourselves to the case where the cosmological constant $\Lambda =0$.}. 

\section{Towards a theory of macroscopic gravity}

The above discussions indicate that the treatments of GR (as it stands) as
either microscopic or macroscopic are problematic. The question is how to
remedy this fundamental shortcoming? One line research has been to attempt
to construct a macroscopic theory, starting with GR as a microscopic theory
of gravity. This is in fact the reverse of the situation that arose in
electromagnetism, where it was proposed by Lorentz that, in addition to the
usual macroscopic theory of Maxwell, there exists a microscopic level of
description of electromagnetic phenomena \cite{Lore:1916}, from which the
macroscopic theory can be derived by an appropriate spacetime averaging
procedure. \\

The formulation of a macroscopic theory of gravity, in the sense of Lorentz,
is, however, much more complicated and requires the following three problems
to be tackled: (i) how to define spacetime averages on a curved spacetime
manifold, (ii) how to average the left hand side of the EFE to establish the
form of the macroscopic field operator and (iii) how to average the
microscopic matter distribution on the right hand side of the EFE in order
to construct a macroscopic model of gravitating matter. \\

To accomplish (i), one requires a generalisation of the spacetime averaging
procedure for flat space, defined in Cartesian coordinates (see, for
example, \cite{Pano-Phil:1962,Rose:1965}). An important feature of such
averages is that they keep the volume of the spacetime regions constant so
as to ensure their applicability to all scales of interest. In this way the
volume may be taken as a free parameter of the averaged theory, with the
corresponding equations valid on any scale. The particular value of this
volume (scale) would then need to be fixed by a model of the matter. In this
connection, it has recently been shown (see \cite
{Zala:1992,Zala:1993,Zala:1996}) that there exist a class of
volume-preserving idempotent averaging operators, which allows the
definition of a covariant spacetime averaging procedure, by generalising the
flat spacetime case and keeping the averaging volume as a free parameter. \\

The resolution of the question (ii), however, turns out to be much more
difficult than the associated question in electrodynamics. The main reason
is that, as distinct from Maxwell-Lorentz equations, the EFE are nonlinear
with the important consequence that the field correlation functions arising
in the process of averaging cannot be defined in terms of the EFE
themselves. This is a direct consequence of the fact that in such settings
the averages of the products of a quantity $Q$ are not equal to the products
of its averages, i.e. 
\begin{equation}
\label{product}\langle Q^m\rangle \langle Q^n\rangle \neq \langle
Q^{m+n}\rangle ,
\end{equation}
where $m$ and $n$ are positive integers and $\langle ..\rangle $ denotes an
average. As a result, averaging in general requires correlation terms which
need to be specified {\it externally} to the theory. In this way the
averaging procedures are {\it non-unique} in at least two ways: firstly due
to the freedom that exists in the choice of the procedure itself and
secondly due to the assumptions necessary to estimate the correlation terms
within each procedure. A number of such procedures have been proposed in the
literature (for a review see \cite{Kras:1996}), but these are mostly
perturbative in nature and only go as far as the second order in
perturbations. Nevertheless, all these schemes have already demonstrated
that any attempt to average out the terms of the second order in the metric
perturbations\footnote{We should point out that in 
perturbative approaches $Q$ is usually taken to be 
a metric perturbation with the property
that $\langle Q \rangle =0$. But the second order term
 (\ref{product}) is already non-trivial, $\langle Q^2 \rangle \ne 0$.} does
require the evaluation of correlation terms which again can only be defined
externally to the theory. Consequently, it seems that the elements of
externality and non-uniqueness are generic properties of all such averaging
procedures\footnote{There are similarities here with the
so called {\it closure problem} \cite{Lesl:1973}, which arises in the 
formulation of the problem
of turbulence, where the moment equations need to be truncated
(closed), by hand, i.e. {\it externally}, which in turn leads to the
non-uniqueness of the procedure.}. \\

A non-perturbative approach recently put forward by Zalaletdinov \cite
{Zala:1992,Zala:1993} (see reviews on the approach in \cite
{Zala:1996,Zala:1994,Zala:1995}) consists of a spacetime averaging of
Cartan's structure equations for the (pseudo)-Riemannian geometry of
spacetime and the EFE. The outcome of this procedure is a set of averaged
equations in the form 
\begin{equation}
\label{eq:aver-Einstein}M_{ab}-\frac 12G_{ab}G^{cd}M_{cd}=-\kappa T_{ab}^{
{\rm (macro)}},
\end{equation}
where $G_{ab}$ is the macroscopic metric, $M_{ab}$ is the Ricci tensor
corresponding to the Riemannian curvature tensor $M^a{}_{bcd}$ and the
macroscopic stress-energy tensor $T_{ab}^{{\rm (macro)}}=T_a^{c{\rm (macro)}
}G_{cb}$ is of the form 
\begin{equation}
\label{eq:macro-energy}T_b^{a{\rm (macro)}}=\langle t_b^{a{\rm (micro)}
}\rangle +C_b^a,
\end{equation}
where $\langle t_b^{a{\rm (micro)}}\rangle $ is the averaged energy-momentum
tensor and $C_b^a$ embodies the field correlation terms. These equations
have been shown to possess the following properties: (i) in the high
frequency limit (up to second order perturbations) they reduce \cite
{Zala:1993} to Isaacson's equations \cite{Isaa:1968} and (ii) in the case
where all correlation functions vanish they reduce \cite{Zala:1996} to the
usual EFE 
\begin{equation}
\label{macro:cosmos}M_{ab}-\frac 12G_{ab}G^{cd}M_{cd}=-\kappa \langle
t_{ab}^{{\rm (micro)}}\rangle ,
\end{equation}
where in cosmological applications, the tensor on the right hand side is
usually taken as a perfect fluid stress-energy tensor. Equations (\ref
{macro:cosmos}) are of the same form as (\ref{efe:ss}), together with $
\langle t_{ab}^{{\rm (micro)}}\rangle =T_{ab}^{{\rm (macro)}}=T_{ab}^{{\rm 
(hydro)}}$ and the identification $M_{ab}=R_{ab}$. An interesting outcome of
this that it makes transparent and precise the implicit assumptions that are
usually made in general relativistic cosmological modelling, namely, that
all correlation functions due to the macroscopic nature of gravity vanish. \\

Finally, point (iii) concerns the question of construction of a model of the
macroscopic gravitating matter in the form 
\begin{equation}
\label{aver:matter}\langle t_{ab}^{{\rm (micro)}}\rangle =T_{ab}^{{\rm 
(hydro)}}. 
\end{equation}
In analogy to electrodynamics, this would require a model of microscopic
matter to be specified, which is a very difficult task with no clear
indications so far as to how to proceed (see an approach in \cite{Szek:1971}
). The usual practice in cosmology, however, is to assume a phenomenological
model for $T_{ab}^{{\rm (hydro)}}$, such as a perfect fluid, the validity of
which remains unclear. \\

To conclude, despite important steps that have been taken towards a
macroscopic extension of GR, major problems remain including the questions
of externality and non-uniqueness that seem to accompany the processes of
averaging, as well as the question of construction of appropriate matter
sources. The task of completing this programme remains paramount, especially
in view of recent cosmological observations which seem to indicate that the
dynamical effects of such correlation functions may be of relevance on the
dynamics of large enough scales. \\

Finally, we should mention that the spacetime averaging of Cartan's
structure equations for the (pseudo)-Riemannian geometry of spacetime seems
to lead naturally to non-Riemannian features \cite{Zala:1992,Zala:1993}.
This is interesting, especially in view of the fact that, as was discussed
in Section 2, attempts at describing the motion of extended bodies also lead
to non-Riemannian frameworks. This seems to indicate that a successful
description of extended gravitational phenomena is likely to involve
non-Riemannian considerations. \\


\section{Conclusions}

By considering a number of applications of GR in real settings, as well its
relation to real observations, we argue that GR, as it stands, is deficient
as a classical theory of gravity. This seems to be true whether GR is
treated as a microscopic or macroscopic theory. In this way, therefore,
there seems to be no scales over which the theory, as it stands, holds
precisely. There is a sense, however, in which the treatment of GR as a
microscopic theory is more consistent, for in such a setting it possesses
all the problems ``typical'' for known classical microscopic theories.\\

We briefly discuss some of the recent attempts to construct a macroscopic
theory of gravity, starting from GR as a microscopic theory, as an attempt
to remedy this fundamental shortcoming. We conclude that such constructions
are likely to include external features, be non-unique and involve
non-Riemannian geometrical frameworks. In this way a full consistent theory
of classical gravity with a build-in length remains to be developed. This is
important, not only as a matter of principle, but especially in view of
recent work which indicate that averaging could have important consequences
for the dynamics of the universe and might, for example, be capable of
resolving the so-called  age problem in cosmology (see an approach in \cite
{Bild-Futa:1991}). \\

From the perspective of our discussion here, it is worthwhile to recall that
the text book successes of GR as a theory of gravity rely on the fact that
the setting usually chosen for the classical tests (the one body problem) is
vacuum, and effectively microscopic in nature. As a result, their good
agreement with observations is no surprise as the errors (due to the
deviations of the Sun from an effective point particle) are likely to be
small. What needs to be done is to extend GR in order to provide a
physically and mathematically adequate framework in order to estimate the
errors involved in settings which are {\it far from microscopicity}, as in
the case of the motion of extended bodies and cosmology. The non-uniqueness
of the macroscopic extensions of GR are only likely to be removable
ultimately in reference with observations in such settings.

Finally, since the main points raised here are also applicable to
alternative theories of gravity, our discussion is also of potential
relevance to testing and assessing the viability of such theories in real
settings.

\section*{Acknowledgments}

RT was supported by PPARC UK Grant number H09454. RZ was supported by a
Royal Society fellowship and would like to thank the School of Mathematical
Sciences for hospitality. We also would like to thank George Ellis for
helpful comments on the manuscript and discussions, and Henk van Elst for
careful reading of the maniscript and comments.

\newpage

\end{document}